\documentclass[a4paper,11pt]{article}

\pdfoutput=1 
\usepackage{cite}
\usepackage{jheppubmod}
\usepackage{enumitem}
\usepackage{amsmath}
\usepackage{physics}
\usepackage{graphicx}
\usepackage{amsfonts} 
\usepackage{amssymb}
\usepackage[dvipsnames, svgnames, x11names ]{xcolor}
\usepackage{color}
\usepackage{braket}
\graphicspath{ {./1 /} }
\usepackage{subcaption}
\usepackage{graphicx}
\usepackage[T1]{fontenc} 
\usepackage{tikz}
\usepackage{physics}
\usepackage{amsmath}
\usepackage{dsfont}

\usepackage{graphicx}
\usepackage{bbold}
\graphicspath{ {./1 /} }

\title{Algebras and their Covariant representations in quantum gravity}

\author{Eyoab Bahiru}
\affiliation{SISSA, International School for Advanced Studies, Via Bonomea 265, 34136 Trieste, Italy}
 \affiliation{INFN, National Institute for Nuclear Physics,  Sezione di Trieste,
Via Valerio 2, 34127 Trieste, Italy}
 \affiliation{ICTP, International Centre for Theoretical Physics, Via
Strada Costiera 11, 34151, Trieste, Italy}
\emailAdd{ebahiru@sissa.it}

\date{\today}

\abstract{ 
We study a physically motivated representation of an algebra of operators in gravitational and non gravitational theories called the covariant representation of an algebra. This is a representation where the symmetries of the operator algebra are implemented unitarily on the Hilbert space. We emphasize the very close similarity of this representation to the crossed product of an algebra. In fact, as an example of (and sometimes identified with) a covariance algebra, the crossed product of an algebra is in one to one correspondence with the covariant representation of the algebra. This will in turn 
illuminate physically what the crossed product algebra is in the context of quantum gravity.  
}

\begin{document}
\maketitle
\section{Introduction}

A more precise understanding of algebras of operators in several contexts in quantum gravity has been given much attention in recent years. Even though studying the algebra of the operators of the semiclassical physics had already proven to be useful, for instance in the background of a black hole, the Tomita Takesaki theory of the algebra of operators was used in the reconstruction of the interior of a black hole\cite{Papadodimas:2012aq}, the renewed interest followed the work of Leutheusser and Liu\cite{Leutheusser:2021frk,Leutheusser:2021qhd}. They identified the algebra of operators of a CFT that is thermally entangled with another CFT above the Hawking-Page temperature to be a type III$_{1}$ von Neumann algebra in the strict large $N$ limit. This identification naturally led them to propose an operator that is associated with an infalling observer and discuss the emergence of time in the eternal black hole background in AdS. This was followed by several works of Witten et. al. \cite{Witten:2021unn,Chandrasekaran:2022cip,Chandrasekaran:2022eqq,Penington:2023dql}, where gravitational interactions are added in some limited fashion to the cases where matter does not gravitationally backreact, where the upshot can be summarized to be the Lorentizan derivation of the generalized entropy of the semiclassical states, among others. This is relevant given the role the entropy plays in the understanding of the black hole information paradox and most of the recent applications of quantum information in quantum gravity. Some other directions this analysis proceeded include understanding subregions in several backgrounds, the de Sitter spacetime and Hilbert space, the role of the observer in de Sitter and cosmology in general, progress towards understanding the presence/ absence of the ensemble averaging for black holes in higher dimensions, background independent description of the perturbative quantum gravity to mention few among many\cite{Witten:2023xze,AliAhmad:2023etg,Jensen:2023yxy,Witten:2023qsv,Gomez:2023wrq,Gomez:2022eui,Schlenker:2022dyo}.

The key role of algebras was noticed even in the early days of quantum field theory in curved spacetime \cite{Wald:1995yp}. The reason for this was the fact that the well understood quantization procedures of a classical theory to a quantum theory faces serious problems when applied in curved spacetimes. In essence, there is no unique quantization of the classical theory, rather one arrives at several unitarily inequivalent Hilbert spaces (see \ref{alge} for more discussion) describing possibly 'inequivalent' quantum theories. This is persistent in particular when the background geometries are open with no particular asymptotics (check \cite{Witten:2021jzq} to see the discussion of this issue in several geometries). The resolution is to consider the algebra of operators in the algebraic quantum field theory sense instead, where such arbitrariness is not present when we pass from the classical theory to the quantum theory. The issue is quite similar to what we face as we move from special relativity, where there is a preferred frame of reference, to general relativity where any frame of reference is equivalent to any other, thus the theory should be described in a frame independent manner. The algebra of operators for the quantum theory already includes the physical content of all the unitarily inequivalent Hilbert space at once and is the correct way to describe the quantum theory. Thus it is not quite naive to imagine the useful role played by the algebra of operators in different limits of quantum gravity. 

This article follows a similar spirit in that, we start with an algebra of operators \emph{then} we consider a particular useful representation of that algebra in different situations. This representation is called the covariant representation \cite{1966CMaPh...3....1D,Borchers1966EnergyAM,Takesaki1967CovariantRO} and it includes the representation of the algebra that acts on a certain Hilbert space. But, in addition the symmetries or the automorphisms of the algebra are also implemented as unitary operators on the Hilbert space. For the obvious reason that symmetries are just changes in the our perspective that should not change experimental results, they should be represented by unitary operators in the Hilbert space and this is the only relevant representation for the algebra for physical systems \cite{Borchers1969OnTI}.   

Coming back to the recent developments in algebras and quantum gravity, a certain construction called the crossed product construction has been used to go beyond the strict $G_{N}\rightarrow 0$ limit. In most cases, this led to the change in the type of the von Neumann algebra from a type III$_{1}$ to type II, where a finite trace and entropy can be defined. But still, it seems that this step is a bit mathematical and not physically clear. On the other hand, the covariant representation of an algebra can be rigorously shown to be in one to one correspondence with what is called a covariance algebra (see appendix \ref{covalg}), of which the crossed product algebra is an example \cite{Takesaki1973DualityFC}. Thus, in terms of the covariant representation, the crossed product type II algebra can be understood in a physically intuitive way, which is one of the goals the article hopes to accomplish. 

We can also imagine a case where the vacuum of the Hilbert space breaks some of the symmetries. These cases will lead to what we call proto quantum gravity Hilbert space which is a Hilbert space not described by a quantum field theory on a curved spacetime even though it arises in the strict $G_{N}\rightarrow 0$ limit. An example of such a Hilbert space was discussed in the \cite{Chandrasekaran:2022eqq} and dubbed proto holographic black hole by the authors.   

In section \ref{sec 2}, we revise the relevant background discussion about algebra of operators with out gravitational backreaciton. In section \ref{sec3}, we discuss what the covariant representation of an algebra is and which algebra precisely we are talking about in the most general cases. Then in the following section, we continue the discussion to proto QG Hilbert spaces and covariant representations for the algebra of operators associated with subregions and an observer's worldline in any spacetime.

\section{No Backreaction : review}\label{sec 2}

\subsection{The strict large N limit}\label{rev}

Holography implies that the semiclassical physics in AdS is emergent from a low lying sector of the boundary CFT. For a CFT containing a large central charge and operators that factorize, called generalized free fields, there is a reorganization of the small number of these degrees of freedom (compared to the central charge) that reproduces the gravitational theory in AdS \cite{El-Showk:2011yvt,Heemskerk:2009pn,Heemskerk:2010ty,Fitzpatrick:2010zm,Penedones:2010ue}. In fact this property of generalized free fields, that they obey an equation of motion in a higher dimension even though they do not satisfy any equation of motion in the CFT was well known even before the formulation of AdS/CFT \cite{Haag1962THEPO}. But generalized free fields are not fully self consistent CFTs by themselves and can only be understood as a small sector of a bigger CFT. This property has been useful in understanding of how gravity encodes information as a quantum error correcting code \cite{Almheiri:2014lwa}, granting the Hilbert space they are acting on, the name the \emph{code subspace}. 

Specializing to $\mathcal{N}=4$ super Yang Mills in $4d$ in the 't Hooft limit, the generalized free fields  will be the single trace operators with thermal or vacuum expectation value subtracted. Their correlators factorize into two point functions in the large $N$ limit where $SU(N)$ is the gauge group. We are interested in studying quantum field theory on the eternal black hole in AdS, which is expected to be dual to two thermally entangled CFTs \cite{Maldacena:2001kr} above the Hawking Page temperature. Thus we consider the Hilbert space $\mathcal{H}= \mathcal{H}_{L}\otimes\mathcal{H}_{R}$, where $\mathcal{H}_{L}$ and $\mathcal{H}_{R}$ are the Hilbert spaces of the left and right CFTs. We denote the set of all bounded and linear operators acting on $\mathcal{H}$ by $\mathcal{B(H)}$. The generalized free fields, $\mathcal{\tilde{A}}_{L/R} \subset \mathcal{B(H)}$, acting either on the left or right side form a von Neumann algebra in the strict large $N$ limit (i.e, $1/N$ is set to zero), once their weak closure is taken. The weak closure is a requirement to include any limit point for a Cauchy sequence of matrix elements, i.e, if for $a_{n}\in \mathcal{\tilde{A}}_{L/R}$, lim$_{n \rightarrow \infty}\bra{\psi}a_{n}\ket{\chi}=\bra{\psi}a\ket{\chi}$, for all $\ket{\psi} $ and $\ket{\chi}$ in $\mathcal{H}$, then $a$ must also be in $\mathcal{\tilde{A}}_{L/R}$. If it had not been for the subtraction of the thermal expectation value in the definition the generalized free fields, this algebra $\mathcal{\tilde{A}}_{L/R}$, would have described the entangled system in a background independent way. But following \cite{Leutheusser:2021frk,Leutheusser:2021qhd}, we will specifically discuss the eternal black hole at a given temperature. What Leutheusser and Liu did was to identify the Hilbert space on the eternal black hole background at some temperature $1/\beta$, $\mathcal{H^{\beta}}_{HH}$, with what is called the GNS Hilbert space built on the thermofield double state of the same temperature, which is distinct from $\mathcal{H}$. 

We consider the thermofield double state at inverse temperature $\beta$,
\begin{equation}\label{innerproduct}
    \ket{\Psi_{\beta}}=\frac{1}{Z_{\beta}} \sum_{i}e^{-\beta E_{i}/2}\ket{E_{i}}_{R}\ket{E_{i}}_{L}
\end{equation}
where $\ket{E_{i}}$ are the energy eigenstates. We can define an inner product on $\mathcal{\tilde{A}}_{R}$ using $\ket{\Psi_{\beta}}$ in the large $N$ limit, in particular,
\begin{equation}
    \begin{split}
 \braket{a|b}= \lim_{N \to \infty} \bra{\Psi_{\beta}}ab\ket{\Psi_{\beta}} \,,
\qquad
\forall a,b \in \mathcal{\tilde{A}}_{R} \,.
\end{split}
\end{equation}
Note that, if $\lim_{N\rightarrow \infty}\bra{\Psi_{\beta}}x^{\dagger}x\ket{\Psi_{\beta}} = 0 $ for $x \in \mathcal{\tilde{A}}_{R}$, then it follows the Schwarz inequality that $\ket{a} \sim \ket{a + x}$. Let $X$ be the set of all operators like $x$, then we have an equivalence class $\mathcal{\tilde{A}}_{R}/X$ with an inner product. We can then interpret this set as a pre-Hilbert space, where after we take the Hilbert space completion, becomes a Hilbert space. To take the the Hilbert space completion means to add all the limit points of a Cauchy sequence of states, $\{\ket{\psi_{n}}\}$ in the pre-Hilbert space. We get the GNS Hilbert space, $\mathcal{H}^{\beta}_{GNS}$, after we take the Hilbert space completion of the above equivalence class. Thus we have\footnote{This is not exactly correct as will be seen in the sections that follow. The state $\mathcal{H^{\beta}}_{HH}$ is built upon, $\ket{\psi_{HH}}$, should also include a wave function for the right Hamiltonian, which is a delta function.},
\begin{equation}
   \mathcal{H^{\beta}}_{HH} \equiv\mathcal{H}^{\beta}_{GNS}
\end{equation}

There is also a representation $\pi_{\beta}$ of the operator algebra $\mathcal{\tilde{A}}_{R}$ acting on $\mathcal{H}^{\beta}_{GNS}$,
\begin{equation}
\begin{split}
    \pi_{\beta}(c)\ket{a}=\ket{ca} \,,
\qquad
\forall c,a \in \mathcal{\tilde{A}}_{R}
\end{split}
\end{equation}
We will denote this representation by $\mathcal{A}^{\beta}_{R,0}$ (from now on we will only write the superscript $\beta$ when it is necessary) and it is identified with the bulk fluctuations on right exterior of the eternal black hole geometry at temperature $1/\beta$.

\subsection{Modifications to $\mathcal{H}_{GNS}$}\label{modhgns}

The first, and perhaps not so severe, modification to the operator algebras and the Hilbert spaces is to include the contribution of the conserved charges acting on the bulk Hilbert spaces $\mathcal{H}_{HH}$ and $\mathcal{H}_{dS}$. The Hilbert space $\mathcal{H}_{dS}$ is constructed like $\mathcal{H}_{HH}$ with the action of the bulk fluctuations but on the background of the Bunch-Davies state of de Sitter. To identify these charges, we look at the symmetry of the 'vacua' of each Hilbert spaces which is seen by an observer on the right exterior. For the eternal black hole the symmetry group observed by each exterior is $G_{HH}=\mathbb{R} \cross (Spin(4) \cross SU(4)_{R})/\mathbb{Z}_{2}$  while for static patch of deSitter, it is $ G_{dS}=\mathbb{R} \cross SO(d-1)$. Let's call the generators $Q^{\alpha} \in \mathbf{g}_{HH}$ and $Q^{\beta}\in \mathbf{g}_{dS}$. One has to be careful not to naively include these generators to the operator algebras on the Hilbert spaces since they act only in the right exterior. Thus, they do not map a smooth Cauchy surface to a smooth Cauchy surface, in particular, they create a singularity at the horizon. The natural resolution is to impose a brick wall boundary condition close to horizon, renormalize the charges before taking the limit. The same can be accomplished by renormalizing the charges by appropriate power of $G_{N}$ as we take it to zero. In the case of the eternal black hole the situation is cleaner since we can renormalize the boundary operators such that the conserved charges in the boundary, dual to the bulk charges, take the form $N^{2}Tr(L)$ for some operator $L$ with no explicit $N$ dependence. Thus subtracting the thermal one point function and multiplying by $1/N$ ( $\sim \sqrt{G_{N}}$ for $\mathcal{N}=4$ super Yang Mills in 4 dimensions) will make them finite in the large $N$ limit. 

Let's call the renormalized charges $q^{\alpha} \in \mathbf{g}$ and their bounded functions will act on the corresponding Hilbert spaces. The wave function associated with these modes will take values in $L^{2}(\mathbf{g})$. Thus, the first extension to the Hilbert spaces will be $\mathcal{H}_{GNS}\otimes L^{2}(\mathbf{g}_{HH})$ and $\mathcal{H}_{dS}\otimes L^{2}(\mathbf{g}_{dS})$ while the algebras become $\mathcal{A}_{R,0}\otimes \mathcal{A}_{gHH}$ and $\mathcal{A}_{dS}\otimes \mathcal{A}_{gdS}$, where $\mathcal{A}_{gHH}$ and $\mathcal{A}_{gdS}$ are algebras of bounded functions of the charges of $\mathbf{g}_{HH}$ and $\mathbf{g}_{dS}$ respectively.

\section{In the presence of gravitational interactions}\label{sec3} 

\subsection{Algebra of operators on general backgrounds}\label{alge}

In this subsection, we discuss what  we mean precisely by the algebra of quantum fields in general spacetimes. In section \ref{rev}, the elements of the algebra in consideration ($\mathcal{\tilde{A}}_{L/R}$) was given in terms of the dual CFT subtracted single trace operators in the large N limit, which form an algebra of generalized free fields. These operators are well defined even at non perturbative level, though they do not close to form an algebra without the addition of operators not available in the large N limit, for instance product of N$^{2}$ single trace operators. Even though we do not aspire to define operators that make sense non perturbtatively for general spacetime (including non holographic) at the moment, we would like to describe what the elements of the algebra of observables, $\mathcal{A}$, is when gravitational interactions are added in some limited fashion. 

We take this algebra to be a slight modification to the algebra of observables of a quantum field theory on a general curved spacetime \cite{Wald:1995yp}. This modification can be a perturbative interaction between the matter field and the graviton or an addition of modes that correspond to large diffieomorphisms. So, they would not be present for a quantum field theory, without gravity.

We start with the classical field theory describing matter fields and metric fluctuations on top of some fixed globally hyperbolic geometry. If we assume that the matter fields satisfy equations of motion with a well defined initial value problem, then we can describe the system in terms of a phase space, $\mathcal{M}$, of a pair of smooth functions $(\Pi,\Phi)$ defined on a Cauchy hypersurface $\Sigma$. This phase space also has a symplectic structure; that means there is a non degenerate, closed two form $\Omega(.,.)$ on $\mathcal{M}$. The states of the classical system correspond to points on $\mathcal{M}$, while functions (or more precisely functionals), $f: \mathcal{M}\rightarrow \mathbb{R}$, are the observables (we take linear functions of $(\Pi,\Phi)$ as the basis for this set of observables). On the other hand, a quantum theory is described by a Hilbert space, $\mathcal{H}$, and self adjoint bounded operators acting on the Hilbert space. Quantization of the classical system is thus the problem of finding a map from the classical phase space $\mathcal{M}$ and functions on $\mathcal{M}$ to a Hilbert space and self adjoint operators acting on the Hilbert space. But since Hilbert spaces with infinite dimensions (we expect this since the phase space is infinite dimensional) are all isomorphic, the physical content of the problem concerns with the map from the linear functions (classical observables) on $\mathcal{M}$ to operators acting on a certain Hilbert space. Canonical quantization provides such a map by requiring Poisson brackets satisfied by functions on $\mathcal{M}$ be mapped to commutators satisfied by the operators. 

In the case where the geometry has a killing vector that is everywhere timelike, there is a simple way to implement this, that also ensures the correct short distance behaviour of the matrix elements of the quantum observables. For definiteness, let's consider a free scalar field with the equation of motion,

\begin{equation}\label{4.2}
    \frac{1}{\sqrt{g}}\partial_{\mu}(\sqrt{g}g^{\mu \nu}\partial_{\nu}\Phi-m^{2}\Phi) = 0
\end{equation} 
where $g_{\mu \nu}$ is the metric.


The metric can be written in a time independent way and we can define a self adjoint Hamiltonian that is bounded from below. A solution of (\ref{4.2}) can be decomposed into a positive and negative frequency modes of complex functions,  
\begin{equation}\label{4.3}
    \Phi(x,t)=\int d\omega dk \; a_{\omega k} f_{\omega k}(x)e^{i\omega t}+a^{\dagger}_{\omega k} f^{*}_{\omega k}(x)e^{-i\omega t}
\end{equation}
 where $\omega >0$. The phase space of the classical system is the space of functions $(\Pi,\Phi)$, with $\Pi=n^{\mu}D_{\mu}\Phi$ where $D_{\mu}$ is the covariant derivative and $n^{\mu}$ is a normal vector to $\Sigma$. Then, mapping Poisson brackets to commutators leads to the canonical commutation condition,
$[a_{\omega k},a^{\dagger}_{\omega^{'}  k^{'}}]=\delta(\omega-\omega^{'})\delta(k-k^{'})$, while the rest of the commutators vanish.

In addition, we can also define an inner product to be the real part of the Klein-Gordon inner product of the positive frequency parts of the solutions corresponding to the initial data\footnote{Check section 4.3 of \cite{Wald:1995yp} for a precise construction.}. Thus, in this case, we can consider the algebra of operators generated by the a finite but arbitrary product of $a$ and $a^{\dagger}$'s and take our algebra of observables for the quantum field theory to be the completion of this algebra with respect to the inner product just described. Similarly, the algebra can be extended to include other matter fields and free gravitons. 

But the above analysis is only for a stationary spacetimes. In particular, we would not have the decomposition of the fields into positive and negative frequency modes (\ref{4.3}) in a general curved spacetime.\footnote{And with it, we lose the nice particle interpretation of the Hilbert space.} Thus to proceed, we go back to the symplectic structure of the phase space and note that the function $f = \Omega(q,.)$, for $q\in \mathcal{M}$ is a linear function on $\mathcal{M}$ since $f(p)=\Omega(q,p) \in \mathbb{R}$ and the symplectic form is non degenerate. Therefore, we can take $\Omega(q,.)$ as the basis for our set of classical observables. In most cases the phase space is defined as the cotangent bundle of the configuration space of the system and thus the symplectic form is taken to be the usual anti-symmetric bilinear map on a contangent bundle. For example, in the case of the free scalar field mentioned above,
\begin{equation}
    \Omega[(\Pi_{1},\Phi_{1}),(\Pi_{2},\Phi_{2})]=\int_{\Sigma}(\Pi_{1}\Phi_{2}-\Pi_{2}\Phi_{1}).
\end{equation}
Following this, we can summarize the Poisson bracket conditions on $(\Pi,\Phi)$ in terms of the symplectic form as
\begin{equation}
    \{\Omega(q_{1},.),\Omega(q_
{2},.)\}=-\Omega(q_{1},q_{2})I
\end{equation}
where $q_{i}=(\Pi_{i},\Phi_{i})$, $i=1,2$. For instance if we take $q_{1}=(\delta_{\epsilon}(x),0)$ and $q_{2}=(0,-\delta_{\epsilon}(y))$ where $\delta
_{\epsilon}$ is a smooth approximation of the delta function such that $\lim_{\epsilon \rightarrow 0}\delta
_{\epsilon}(x)=\delta(x)$, then in the limit $\epsilon \rightarrow 0$, $\Omega(q_{1},(\Pi,\Phi))=\Phi(x)$ and $\Omega(q_{2},(\Pi,\Phi))=\Pi(y)$ and we find,
\begin{equation}
    \{\Phi(x),\Pi(y)\}=\delta(x-y).
\end{equation}

Since $\Omega(q,.)$ is an observable, we can promote it into self adjoint operator so that 
\begin{equation}\label{elementofa}
    [\bar{\Omega}(q_{1},.),\bar{\Omega}(q_
{2},.)]=-i \Omega(q_{1},q_{2})I
\end{equation}
is satisfied. The bounded functions\footnote{A basis for the bounded functions of the operators can be taken to be the unitary operators generated by $\bar{\Omega}(q_{1},.)$. The self adjointness and the commutation condition on the operators translates to the unitary operators satisfying thw Weyl relation.} of these operators form an algebra\footnote{This algebra is called a $^{*}-$algebra \cite{Haag:1992hx}} and we can also define an inner product from the symplectic form so that\cite{Wald:1995yp},
\begin{equation}\label{innpro}
    \langle q_{1},q_{1}\rangle=\frac{1}{2}\text{sup}_{q_{2}\neq 0}\frac{|\Omega(q_{1},q_{2})|^{2}}{\langle q_{2},q_{2}\rangle}.
\end{equation}
Taking the completion of the algebra with respect to this inner product, we get the algebra that corresponds to a quantum field theory on the curved spacetime.\footnote{If there is a sequence of operators $\{a_{n}\}$ and if the completion of the algebra is taken such that the operator $a$ is in included in the algebra when, for any $\psi$ in the Hilbert space, $a\psi=\lim_{n\rightarrow \infty}a_{n}\psi$, then the algebra is what is called a C$^{*}$ algebra. If we rather take the completion of the $^{*}-$algebra by including the limit points of the expectation values of the operators, we will get a von Neumann algebra.} It should be noted that there is a wide class of inner products that satisfy the above equation and that there is arbitrariness in the definition of the inner product, but this will not create a problem since the resulting algebras are all isomorphic as abstract algebras\cite{Wald:1995yp}. However, this becomes a problem when we construct a Hilbert space. 

To describe the algebra of observables in the presence of perturbative gravitational interactions, we will consider the same algebra but now we can also take linear combination of the operators with powers of $G_{N}$ as coefficients. In the case where we just include a large diffeomorphism mode, we form the algebra associated with it from the unitaries constructed from the generator of the diffeomorphism. The algebra includes bounded functions of the generator and any bounded function can be expressed in terms of the unitaries. We will reformulate the latter case in terms of the covariant representation of the algebra of the paragraph above, in the sections that follow.

Coming back to the construction of the Hilbert space, we will first take the Hilbert space completion of $\mathcal{M}$ with respect to a given inner product, (\ref{innpro}). We then complexify the manifold using the antisymmetric two form $\Omega$. This complex manifold, $\mathcal{\bar{M}}$, can be used to create the Fock space of states, which is given by the direct sum of the vacuum ($\mathbb{C}$), one particle space, two particle space and so on; where the n-particle space is given by the symmetric (bosonic) or antisymmetric (fermionic) tensor product of n copy of $\mathcal{\bar{M}}$. But as mentioned before, the condition (\ref{innpro}) does not uniquely determine an inner product and the different inner products in general lead to different Hilbert spaces. Naturally, it is reasonable to take the direct sum of all the Hilbert spaces and consider a bigger Hilbert space (this Hilbert space is the one studied by von Neumann et. al. \cite{10.2307/1969463,10.2307/1968823} as infinite tensor product Hilbert spaces, where in the same papers shown that they are equivalent to the infinite direct sum Hilbert spaces). But there are a couple of issues with this bigger Hilbert space\cite{Streater:1989vi,Bahiru:2022mwh}, first it is a Hilbert space with unaccountably infinite dimensions. Second, the individual Hilbert spaces are what are called unitarily inequivalent and they correspond to different superselection sectors, in the sense that states in different sectors will not form a coherent superposition and a superposition only results in a mixed state and any dynamics, implemented by a unitary evolution will not evolve states in one sector to a different sector. Hoping to avoid this non-separable Hilbert space and because of the fact that each sector can be treated completely independently, studying the individual sectors only is generally expected to be enough to describe a physical system\cite{Streater:1989vi}. But again a question arises as to which of the sectors we should choose, to correspond to the quantization of the classical system. 

In the case where the geometry has a timelike Killing vector that is globally defined, there is a canonical choice of inner product, in particular a generalization of the inner product mentioned for the free scalar field. This is also directly related with the fact that there is a canonical choice of the vacuum state. In addition to stationary spacetimes, it was also shown that \cite{Witten:2021jzq} for asymptotically AdS (with boundary conditions as that of AdS/CFT), asymptotically flat spacetimes (with theories with a mass gap) and compact spacetimes, there is a canonical choice of vacuum and a natural choice of Hilbert space. For a more general open spacetimes though, it is expected that there is indeed not a canonical inner product (or choice of 'vacuum' state) and an algebraic treatment of the theory is necessary as it includes the content of all the unitarily inequivalent Hilbert spaces in a systematic manner, as was discussed in the introduction\footnote{Additional condition on the states is making sure that they reproduce the correct short distance behaviour for the operators. This is analyzed by checking that the UV behaviour of the two point functions has the correct singularity. This condition goes by the name, the Hadamard condition. Check \cite{Wald:1995yp} for more careful discussion.}. 

This algebra is the same for any spacetime in the sense that it is generated by the (\ref{elementofa}). But the elements differ for the specific choice of geometry since the symplectic form of the classical theory will in general be different. For instance, for flat spacetimes, the operators (\ref{elementofa}) can be shown to be the sum of the annihilation and creation operator \cite{Wald:1995yp}. On the other hand, the fact that even for the same classical theory, there are several unitraily inequivalent Hilbert space will lead to several representations ($\pi$) of the algebra on the Hilbert spaces. These representations can be constructed following the GNS construction (see \ref{rev}) on a certain cyclic state in the Hilbert space. In the following sections, we discuss the covariant representation of this algebra.      

\subsection{Covariant representation of the algebra}\label{covrepalg}

One of the central principles of algebraic approach towards understanding of quantum field theory (and statistical physics) has been that physical content is actually algebraic and does not depend on the representation. Still, there are certain physically sensible and useful representations. In particular, if $\mathcal{A}$ is a von Neumann algebra with an automorphism group $G$, then the most interesting representation (and probably the only relevant representation\cite{Borchers1969OnTI,Borchers1966EnergyAM}) for physics is one where the automorphism/symmetry group is implemented by an action of a unitary operator on the Hilbert space (the reason for considering unitary operators is that the symmetry group is expected to preserve transition probabilities and this is true if they are represented by a unitary operator). This is to say that one concentrates on a set of special states which are related to each other by a unitary action of the automorphism group. The automorphism group can be one associated with space-time translations or some internal symmetry of the system in consideration. Such representation is called a covariant representation of the algebra $\mathcal{A}$\cite{1966CMaPh...3....1D,Borchers1966EnergyAM,Takesaki1967CovariantRO,Borchers1969OnTI}. 

More precisely, a covariant representation $(\pi, U)$ of $\mathcal{A}$ is pair of a non-degenerate representation of $\mathcal{A}$, $\pi : \mathcal{A} \rightarrow \mathcal{B(H)}$ and  a unitary representation of the automorphism group $G$, $U: G \rightarrow \mathcal{U(H)}$, where $\mathcal{B(H)}$ is a set of bounded operators acting on $\mathcal{H}$ and $\mathcal{U(H)}$ is a set of unitary operators acting on $\mathcal{H}$, such that;
\begin{equation}\label{covrep}
\begin{split}
U(g)^{\dagger}\pi(a)U(g)=\pi(\alpha_{g}(a))\,,\qquad \text{ for all } a \in \mathcal{A}
\end{split}    
\end{equation}
where $\alpha : G \rightarrow \text{Auto} \mathcal{A}$ and $\text{Auto} \mathcal{A}$ is the automorphism group of $\mathcal{A}$. We will write the unitary representations of the automorphism group as $U(g)$.

Thus, given an algebra of operators of a quantum system, gravitational or not, \emph{we should ask what is the Hilbert space where there is a non degenerate representation of the algebra. In addition, there should also be a unitary representation of the automorphism group of the algebra acting on the same Hilbert space such that \ref{covrep} is satisfied.}

For instance, in the context of section \ref{rev}, once we are given an algebra of operators $\tilde{A}_{R}$ it is then physically necessary to ask what the covariant representation of the algebra is.  The first guess at the relevant Hilbert space may be $\mathcal{H}_{HH}$, as one is interested in a quantum field theory on the black hole or deSitter background. But, since this Hilbert space does not carry a unitary representation of the automorphism group of $\tilde{A}_{R}$, one should consider an extension of it, say $\mathcal{H}=\mathcal{H}_{HH} \otimes L^{2}(\mathbf{g})$. Then we take the representation, $\pi$ to be the representation furnished on the GNS Hilbert space, $\pi_{\beta}$ (see section \ref{rev}) and identity on the $L^{2}(\mathbf{g})$ factor. The unitary representation of the automorphism group on the other hand acts as identity on $\mathcal{H}_{HH}$ but acts by mulitplication on the $L^{2}(\mathbf{g})$ factor. We will discuss in section \ref{infallobsewmass} another representation of $\tilde{A}_{R}$ that is also covariant.

Even though, as stated before, this statement is to be applied to either gravitational or non gravitational systems, there are quite interesting features that arise in gravitational setting and also in non gravitational systems when we think of them as limits of gravitational systems. By a gravitational system we mean that a system where gravitational modes are present and interact with the rest of matter fields perturbatively (AdS/CFT will be the special case where 'we know what we mean' non-perturbatively).

Such a system in general will be described by an action of the form,
\begin{equation}\label{act}
    S=\frac{1}{G_{N}}\int dx^{d} \sqrt{-g}(R+ ...) 
\end{equation}
where the $ ... $ represents the matter part of the action with minimal coupling to gravity and a boundary Gibbons-Hawking-Yorks term. A non gravitational system will be described by the strict $G_{N}\rightarrow 0$ limit of the above action. We assume the full geometry to be asymptotically flat or AdS but we have not specified the background geometry with respect to which we take this limit, $g_{\mu\nu}=g^{0}_{\mu\nu}+\sqrt{G_{N}}h_{\mu\nu}$. We can consider the matter fluctuations (and free graviton) acting on the full spacetime or a subregion and take the algebra of these operators as the starting point. The physically relevant question will be what is the covariant representation of this algebra, i.e, the appropriate Hilbert space and representation for the algebra, where the automorphisms of act unitarily?

To answer this question, 1) we must know what the symmetries (automorphisms) of the algebra of operators are; 2) we must find the Hilbert space where they act unitarily. As we consider the $G_{N}\rightarrow 0$ limit, we note the following; the associated charges for the automorphisms can be obtained from the action (\ref{act}). Thus the generators for the autmorphisms (rather for the unitary representation of the automorphisms) will in general also have the following form\footnote{In general there are more than one generators for the representation of $G$},
\begin{equation}\label{chr}
\begin{split}
    Q=\frac{1}{G_{N}}\int dx^{d} f(x)\,,\qquad\text{ and } [Q,\pi(a)]=O(1)
\end{split}
\end{equation}
where $f(x)$ is a function of the background metric, the metric fluctuations, the matter fields and their derivatives.

As it is the case for the action, we have to subtract the contribution from the background space-time and redefine the charge $Q$ to get a possibly finite result. But, as we take the limit $G_{N}\rightarrow 0$, this only gives finite expectation value for $Q$ except in the cases where the background spacetime is flat or pure AdS. (There are also other exceptions, specifically states with $O(1)$ variance for the charges. We will defer the discussion of these states to section \ref{infallobsewmass}.) In the other most interesting examples, black holes or semiclassical geometries of coherent states \cite{Skenderis:2008dg,Botta-Cantcheff:2015sav,Marolf:2017kvq,Belin:2018fxe}, (\ref{chr}) diverges specifically as a result of the $G_{N}\rightarrow 0$ limit. In particular such semiclassical states will have $O(1/G_{N})$ variance in the charges\cite{Bahiru:2023zlc}.\footnote{For each Hilbert space we assume the existence of a Hadamrad vacuum with respect to which the n-point functions will satisfy the Wick contraction and factorize to 2 point functions.}

Thus in the strict $G_{N}\rightarrow 0$ limit in particular, the charge that is well defined is,
\begin{equation}\label{renchr}
    q= \sqrt{G_{N}}Q.
\end{equation}
and together with (\ref{chr}), we have $[q,\pi(a)]=0$. This implies that in most interesting cases (except in the cases where the background metric is flat or pure AdS) where the renomarlization (\ref{renchr}) is necessary, the automorphism generators will not produce $O(1)$ transformation, in fact they will be central to the representation of the operator algebra $\mathcal{A}$. The consequence of this fact is that the condition (\ref{covrep}) for the covariant representation  will be trivial in the strict large $N$ limit. 

Thus the covariant representation of the algebra $\mathcal{A}$ with a symmetry group $G$ ( $\mathbf{g}$ being the corresponding lie algebra) will be $(\pi, U)$ acting on a Hilbert space $\mathcal{H}=\mathcal{H}_{bulk}\otimes L^{2}(\mathbf{g})$ where $\pi(a)$ will act on $\mathcal{H}_{bulk}$ creating bulk fluctuations and acts as identity on $L^{2}(\mathbf{g})$ while $U(g)=e^{iq^{\alpha}\nu_{\alpha}}$ acts as identity on $\mathcal{H}_{bulk}$ and by multiplication on $L^{2}(\mathbf{g})$. In addition, $U(g)$ will commute with the bulk fluctuations $\pi(a)$. But note that when the background geometry is flat or pure AdS, $U(g)$ will have non-trivial action on $\pi(a)$ given by (\ref{covrep}). 

This property, that for general gravitational systems in the strict $G_{N}$ going to zero limit the automorphism generators do not generate transformations, may look a little unattractive. Let's divide the symmetry groups into the compact and non compact components. Even though the non compact group be could more general and even non abelian, we specifically treat time translations as the non compact group in part because time translation is usually one of the symmetry generators for $\mathcal{A}$ in part because we want to make contact with the cross product construction of \cite{Witten:2021unn,Chandrasekaran:2022cip} (we defer the discussion on the compact subgroup to section \ref{infallobsewmass}). The above argument implies that time development generator commutes with operators in the algebra. \emph{How would we then implement time translations for our operators?}

Keeping the above question in mind, let's consider the modification of the covariant representation $(\pi,U)$ when gravitational interactions are added perturbatively. Looking at representation $\pi$, we can construct operators with coefficients in the ring of power series in $\sqrt{G_{N}}$. On the other hand, the unitary representations of the symmetry group will transform as follows; concentrating on time translations, as we back away from the strict $G_{N}$ going to zero limit, since $\sqrt{G_{N}}Q$ generates transformations of $O(G_{N})$, the most natural modification to (\ref{renchr}) would be 
\begin{equation}
    \sqrt{G_{N}}Q_{t}=q_{t}-\frac{\sqrt{G_{N}}}{\beta}\text{log}\Delta.
\end{equation}
For asymptotically AdS eternal black hole spacetime, the above modifications can be showed to be enough to all orders in perturbation theory \cite{Witten:2021unn}, "otherwise it is valid to first order in perturbation theory." The unitary representation for time translation symmetry acting on the Hilbert space $\mathcal{H}_{bulk}\otimes L^{2}(\mathbf{g})$ will thus be $U(g_{t})=e^{i(q_{t}-\sqrt{G_{N}}/\beta\;\text{log}\Delta)\nu}$.

This looks quite similar to the cross product construction but our starting point is different and the crossed product construction is in fact distinct from it. The crossed product algebra is a special kind of what is called a covariance algebra (see appendix \ref{covalg}). It was shown in \cite{1966CMaPh...3....1D} that the covariant representation of an algebra $\mathcal{A}$ and the representations of the covariance algebra associated with $\mathcal{A}$ are in one to one correspondence.

Coming back to the question we asked above, the bulk fluctuations now transform non trivially under the action of $U(g)$, following the equation (\ref{covrep}). Still, since we are working in perturbation theory, the transformation is infinitesimally small. To produce an $O(1)$ time translations, we propose (inspired by \cite{Leutheusser:2021qhd}) $U(s)$, the half sided modular translation\cite{Leutheusser:2021frk,Leutheusser:2021qhd,eyoab}, as the appropriate operator whenever it is possible to define it, particularly because its generator has bounded spectrum. Note that for the other candidate for the generator of a translation operator, $-$log$\Delta$, the spectrum is not bounded from below, in general.

\section{Gravitational interaction: continued}\label{infallobsewmass}

\subsection{Proto-QG Hilbert space}

In the previous section, we introduced the covariaint representation of an algebra. The algebra of our interest was the one discussed in subsection \ref{alge}. Up till now we have focused mainly on the algebra where perturbative corrections are added, controlled by $G_{N}$, to the algebra of operators for a quantum field theory on a curved spacetime. Now, we discuss the case where no perturbative correction is added to the QFT operators, rather we consider a 'background' state that has $O(1)$ variance in the charges. Thus, we consider a Hilbert space built on top of such a state with bulk fluctuations and where the symmetries of the spacetime are implemented unitarily. As we will see, such a representation corresponds to a system where gravitational modes that are associated with large diffeomorphism is added to the description. Since this representation is not the naively considered Hilbert space of a QFT on a fixed background and yet it arises from the strict $G_{N}\rightarrow0$ limit of perturbative quantum gravity, we call it proto-quantum gravity (proto-QG) Hilbert space.

Let's consider the covariant representation of the algebra $\mathcal{A}$ on flat or pure AdS spacetimes. Then $\pi(a)$ will be operators that create bulk fluctuations on the flat or AdS spacetime. The unitary representations of the symmetries will be constructed as discussed in subsection \ref{covrepalg}. The only difference being the additional normalization of the charges, multiplication by $\sqrt{G_{N}}$, is no longer necessary. That is, $Q$ is already well defined in the limit $G_{N}\rightarrow0$, and $[Q,\pi(a)]$ is nonzero $O(1)$ number. Thus, the covariant representation $(\pi,U)$ where $U=e^{iQ\nu}$ will act on $\mathcal{H}_{bulk}\otimes L^{2}(\mathbf{g})$. The operators $\pi(a)$ will act on $\mathcal{H}_{bulk}$ creating bulk fluctuations while acting as identity on $L^{2}(\mathbf{g})$ and $U(\mathbf{g})$ acts on $\mathcal{H}_{bulk}$ as identity while acting on $L^{2}(\mathbf{g})$ by multiplication. The fact that $Q$, for gravitational theories in general, is a boundary quantity together with $[Q,\pi(a)]$ shows that the operators $\pi(a)$ are dressed with respect to the boundary of the spacetime and alludes to the gravitational nature of the Hilbert space even in this $G_{N}=0$ limit. Note that the one to one correspondence with the cross product algebra implies that this algebra is a type II$_{\infty}$ algebra. The cross product algebra in Minkowski and pure AdS was discussed in \cite{AliAhmad:2023etg}.  


The gravitational nature of the Hilbert space is even more elaborated in the example of the GNS Hilbert space of the microcanonical thermofield double state \cite{Chandrasekaran:2022eqq}. We consider a new covariant represenation of the same algebra of operators, $\mathcal{\tilde{A}}_{R}$ introduced in section \ref{rev}. Rather than using the GNS Hilbert space of the thermofield double, we consider the state,
\begin{equation}
    \ket{\Phi_{\beta}}=\frac{1}{e^{S(E_{0})}} \sum_{i}e^{-\beta (E_{i}-E_{0})/2}f(E_{i}-E_{0})\ket{E_{i}}_{R}\ket{E_{i}}_{L}
\end{equation}
where, $\int dx\;|f(x)|^{2}=1$ and define an inner product as in (\ref{innerproduct}),
\begin{equation}
     \begin{split}
 \braket{a|b}= \lim_{N \to \infty} \bra{\Phi_{\beta}}ab\ket{\Phi_{\beta}} \,,
\qquad
\forall a,b \in \mathcal{\tilde{A}}_{R} \,.
\end{split}
\end{equation}
If $Y$ is the ideal of the algebra with respect to this inner product, we can define the equivalence class, $\mathcal{\tilde{A}}_{R}/Y$ and complete it to a Hilbert space, $H_{\Phi}$. Thus, we take the non degenerate representation $\pi_{\Phi}$ acting on $H_{\Phi}$ as,
\begin{equation}
    \begin{split}
    \pi_{\beta}(c)\ket{a}=\ket{ca} \,,
\qquad
\forall c,a \in \mathcal{\tilde{A}}_{R}
\end{split}
\end{equation}
But we notice that the symmetry generator of the algebra $\mathcal{\tilde{A}}_{R}$ (we only consider the non-compact subgroup for the moment) is also present in this Hilbert space and is well defined in the large N limit. In particular, it the Hamiltonian of the right system. This is in contrast to the Hartle-Hawking state where this observable is not present and we had to consider an extension of the Hilbert space built on $\ket{\psi_{HH}}$ (by the action of the bulk fluctuations) to get a covariant representation. Even the canonical thermofield double Hilbert space includes a covariant representation, i.e, no extension is needed. The only issue with the canonical thermofield double is that the generator has to be normalized twice to be well defined
in the large N limit. The obvious difference between the canonical or microcanonical thermofield double and the Hartle Hawking state is that, the former are well defined even in non-perturbtive theory of quantum gravity. This again points to the intrinsically quantum gravitational nature the the covariant representation of an algebra.

Since the subtracted generator, $Q$ of the symmetry is already included in the Hilbert space and well defined large N limit, we take $U(g_{t})=e^{iQ\nu}$ as the unitary representation of the symmetry group, where it acts as.
\begin{equation}
    Q=q-\frac{1}{\beta} \text{log}\Delta_{\Psi}.
\end{equation}
This is because it translates the operators $\pi_{\Phi}(a)$ but differs from the modular Hamiltonian by a central operator, $q$. Thus the representation $(\pi_{\Phi},U)$ acting on the Hilbert space $H_{\Phi}$ will be a covariant representation of $\mathcal{\tilde{A}}_{R}$ and it is an example where the variance of the generator is $O(1)$.

The states in the Hilbert space correspond to different semiclassical states in general even though we have set $G_{N}$ to zero. The reason is that the solution $\ket{\Phi_{\beta}}$ breaks the symmetry of the theory. In particular, while the symmetry (corresponding to time translations) of the full theory is $\mathbb{R}_{R}\otimes \mathbb{R}_{L}$ generated by the left and right Hamiltonians, the solution is only symmetric under the diagonal sub group $\mathbb{R}$, generated by the difference of the two Hamiltonians. This implies that there is a moduli space of classical solutions given again by $\mathbb{R}$. Since $U$ only acts on the right boundary, states related by the action of $U(g_{t})$ are associated with different classical solutions on the moduli space. The action of $\pi_{\Phi}(a)$ on any of the classical solutions
would correspond to a given 'Hilbert space' of QFT on a curved spacetime. That is to say that, excluding $U$, one of these states acted upon by the bulk fluctuations $\pi_{\Phi}(a)$ is what we expect QFT on a curved spacetime describes. Note that these Hilbert spaces are not the unitraily inequivalent Hilbert spaces described in section \ref{alge}, rather they more similar to the sectors corresponding to different charge sectors of a gauge theory. This sectors in a gauge theory are related by a large gauge transformation that does not vanish at infinity. Similarly, the transformation by $Q$ is a diffeomorphism that actually does not vanish in the boundary. This mode parameterizes the timeshift between the left and the right boundary. Thus, we can think of the algebra of operators $(\pi_{\Phi},U)$ as an addition of a gravitational mode to the algebra of operators describing bulk fluctuations at $G_{N}=0$. 

A general state will be a superposition of states with different classical background, thus does not have a semiclassical bulk dual geometry. But, following \cite{Chandrasekaran:2022eqq}, we call a semiclassical state a state with a very large variance in $Q$ (very small variance in the dual mode). The ideal case would be if the variance of $Q$ is $O(1/\sqrt{G_{N}})$, where we get the canonical thermofield double and so a fixed background geometry. 

This feature of the spontaneous breaking of the symmetries can be generalized to any state. Starting with an algebra of operator $\mathcal{A}$ with a symmetry group $G$, we can choose a classical solution in the theory that breaks the symmetry to $G_{s}$. This state can be a coherent state like a collection of galaxies, a black hole or a collapsing geometry. Taking this state as the background geometry we can construct the GNS Hilbert space and describe the quantum field theory on this fixed background (with all the subtleties described in section \ref{alge}). Because the 'vacuum' breaks the symmetry $G$, there is a moduli space of classical solutions parameterized by the quotient $G/G_{s}$. But if we look at this state in the limit $G_{N}\rightarrow0$ of perturbative quantum gravity, charges of the symmetry group $G/G_{s}$ must have variance of $O(1/G_{N})$, since the strict $G_{N}\rightarrow0$ limit is a quantum field theory on a fixed spacetime while the action of the the generators of $G/G_{s}$ map a classical solution of the theory to a different one (see also section \ref{covrepalg}). Thus these generators will have to be normalized (multiplied by $\sqrt{G_{N}}$) to be well defined. On the other hand, if we rather take an $O(1)$ superposition of states that correspond to different classical solutions on the moduli space as a starting point for the covariant representation of $\mathcal{A}$, we have mitigated the diverging variance of the charges, at the cost of not having a well defined semiclassical spacetime. 

As mentioned before there will be a Hilbert space for each element of the moduli group $G/G_{s}$ that corresponds to the classical solution. Thus we have a fiber of Hilbert spaces $\mathcal{F}$ on the base space $G/G_{s}$. The Hilbert space the covariant representation $(\pi,U)$ will act on will be $L^{2}$ sections of the fiber $\mathcal{F}\rightarrow G/G_{s}$. This is the proto-QG Hilbert space for general states and general symmetry group $G$ that can also be compact. For a symmetry group that is non compact, if the algebra generated by $\pi(a)$ for $a\in \mathcal{A}$ is a type III$_{1}$
algebra, then the one to one correspondence with the crossed product algebra implies that the algebra generated by $\pi(a)$ and $U$ will be a type II$_{\infty}$ algebra, with a well defined trace and entropy.

\subsection{Subregions and the observer}

The last comment we would like to add is concerning sub regions. The naive algebra of operators we associate to subregions are not so well defined in the perturbative theory of quantum gravity since the spacetime is fluctuating, we can not define a fixed subregion with respect to which we define the algebra, all the while the spactime fluctuation itself is expected to be included in the algebra of observables (check \cite{Witten:2023qsv,Jensen:2023yxy,Aguilar-Gutierrez:2023odp} for discussions of algebra of operators associated with subregions in perturbative quantum gravity). But we have a well defined notion of subregions in the QFT on a curved spacetime limit and we can define a covariant representation for this algebra associated with a subregion. The symmetry of a sub region has to preserve the causal diamond of the subregion and one such symmetry is the one that looks like the Lorentz boost symmetry close to the horizon, that acts both on the subregion and its complement. But note that the boost generator that only acts on the subregion is not a well defined operator. In fact we have to do a brick-wall regularization to the operator by demanding a Dirichlet boundary conditions on the fields at some proper distance $\epsilon$ from the horizon. In the limit $\epsilon \rightarrow0$, we will have a divergence similar to the $G_{N}\rightarrow0$ divergence of the symmetry generators discussed in \ref{covrepalg}. This brick-wall regularization is clear and unambiguous (up to the Weyl anomaly present in even dimensions) in the cases where the theory has a holographic dual, where the renomlization corresponds to a renomalization of Euclidean hyperbolic manifolds in the bulk dual theory,\cite{Bahiru:2022mwh} (assuming large rank for the gauge group of the boundary theory). Then the algebra generated $\pi(a)$ and $U$ will be a type II$_{\infty}$ algebra.  

Recently, the role of an observer in connection to the algebra of observables in subregions and compact spacetimes like deSitter has been given some attention\cite{Witten:2023qsv,Jensen:2023yxy}. The connection between the two is given by the timelike tube theorem (\cite{Witten:2023qsv,Strohmaier:2023opz}) which states that for the so called 'complete' theories, where all the possible  electrically and magnetically charged objects (particles, strings, ...) that can couple to the gauge fields (and higher form gauge fields) in the theory are also present, that the algebra of quantum fields smeared along a timelike world line is same as the algebra of quantum fields that are causally accessible to the world line. In other words, if we pick two points, $p$ and $q$, one to the future of the other along a timelike world line, the algebra of operators that we can construct by smearing the fields in this segment of the worldline is the same as the algebra of operators that are defined in the causal diamond defined by the points $p$ and $q$. The theorem is proved for quantum field theories in flat and curved spacetimes, \cite{Strohmaier:2023hhy}. 

The fact that the static patch of deSitter can be understood as the region that is causally accessible to an observer and that algebra of operators along the worldline of the observer is the same as that of the static patch (in QFT on a curved spacetime limit) motivated including an observer in the study the algebra of operators in the static patch in perturbative quantum gravity, where since the spacetime has compact spacelike slices the symmetries of the spacetime are to be treated as gravitational constraints. The algebra of operators is defined in the presence of an observer at one of the boundaries of the Penrose diagram of deSitter, which would otherwise be trivial (just c-numbers). 

In our situations, even though it is not shown that the timelike tube theorem is still present in perturbative quantum gravity (as far as the author is aware), the fact that it holds in the QFT limit is enough motivation to study the covariant representation of the algebra of operators along a timelike worldline. Thus we consider an observer in a certain spacetime and the algebra of observables, $\mathcal{A}_{ob}$ constructed from the quantum fields smeared along its worldline.
\begin{equation}
    \Phi=\int d\tau f(\tau)\phi(\tau,x)
\end{equation}

These operators are well defined operators that act on the code subspace of states, that map normalized states to normalizable states.\footnote{A general smearing of a quantum field would not map normalized states to normalized states because of the OPE singularities of the quantum fields.} Since in quantum filed theory the algebra of this operators is the same as the algebra of operators in the subregion that is causally accessible to the worldline, the symmetries of this subregion are also the symmetries of the algebra along the world line. Thus the covariant representation for the algebra of operators along a timelike world line includes a representation of the algebra of operators $\pi(\Phi)$\footnote{This representation is a representation of operators of the theory on the specific code subspace that includes the observer.} and unitaries, $U$, generated by the symmetry generators of the subregion (let's call them generically $Q$) associated to the worldline by timelike tube theorem. Note that this symmetry generators will preserve the worldline but since we assumed the existence of an observer, we will need to add operators that are associated with the observer itself. Following a minimal model for the observer as a clock with a given rest mass, $m$, the only operator that needs to be modified is the time translation operator.
If $Q_{t}$ is one of the symmetry generators that generate time translation along the worldline of the observer, the actual operator that acts on the observer, generating time translations, has to be modified to $Q_{t}+q_{ob}$ so that its spectrum is bounded from below by $m$. With the appropriate renormalization of the symmetry generators, as discussed at the beginning of this subsection, the covariant representation of the algebra of operators for an observer will be $\pi(\Phi)$ and $U$, generated by the symmetry generators where the time translation generator is modified as $Q_{t}+q_{ob}$. 

Following the timelike tube theorem, the operators $\pi(\Phi)$ associated with a time segment on the worldline (or for an observer living at the boundary of one of the static patches) is actually a type III$_{1}$ von Neumann algebra. On the other, the algebra of operators generated by $\pi(\Phi)$ and $U$ will be type II. Simply renormalizing the generators would result in a type II$_{\infty}$ algebra but with the presence of an observer, the algebra will be type II$_{1}$, because of the bounded spectrum of the time translation generator. 

Notice that the reason for the algebra associated, for example with the deSitter static patch, is a type II$_{1}$ algebra is different from the above argument. In \cite{Chandrasekaran:2022cip}, with the addition of the observer to the deSitter patch the algebra is modified from $\mathcal{A}_{0}$ (algebra of quantum fluctuations in the patch) to $\mathcal{A}_{0}\otimes B(L^{2}(\mathbb{R}))$, where $B(L^{2}(\mathbb{R}))$ is the set of bounded functions of the observer's Hamiltonian, $h_{ob}\geq m$. The algebra of operators associated with perturbative QG is the subalgebra that is invariant under the action of the full symmetry generator $h+h_{ob}$, where $h$ is the symmetry generator of deSitter that preserve the static patches\footnote{the operator $h$ is in fact the modular Hamiltonian of $\mathcal{A}_{0}$.}. This subalgebra will be composed of $h_{ob}$ and 
\begin{equation}\label{opppp}
    e^{iph}a_{0}e^{-iph},\,\qquad \forall a_{0} \in \mathcal{A}_{0},
\end{equation}
if $p$ is the conjugate operator to $h_{ob}$. This set of operators is a type II$_{1}$ algebra by Takesaki duality.

To relate this algebra to the covariant representation given above, we use the timelike tube theorem and claim that $\mathcal{A}_{0}$ is the same as the algebra generated by $\pi(\Phi)$'s and conjugating the set of operators $\{e^{iph}a_{0}e^{-iph}, h_{ob}\}$ by $e^{-iph}$ would give the covariant representation of $\mathcal{A}_{ob}$ and the condition $h_{ob}\geq m$ transforms to $h+h_{ob}\geq m$, which is the condition satisfied by $Q_{t}+q_{ob}$ for the covariant representation.

Thus we find that the covariant representation of $\mathcal{A}_{ob}$ is the same as the algebra of operators for the deSitter static patch in perturbative quantum gravity. The same can also be said for a subregion, which by the timelike tube theorem is associated to a proper time segment of the observer's worldline. But there is a subtlety here in that the subregion is not well defined in perturbative QG because the spacetime fluctuates with fluctuations of $O(\sqrt{G_{N}})$, in the same sense there is uncertainty of $O(\sqrt{G_{N}})$ in specifying a point in time along the worldline, and so there is uncertainty in defining the segment.




\acknowledgments
I am grateful to K. Papadodimas for useful discussions. I want to thank the CERN-TH for their support and hospitality during the preparation of this paper. The research is also partially supported by the Eramsus+ Trainee-ship program and INFN Iniziativa Specifica String Theory and Fundamental Interactions.

\appendix

\section{Covariance algebra} \label{covalg}

We will now introduce the definition of the covariance algebra and only state some of its properties. We refer the reader to \cite{1966CMaPh...3....1D} for complete discussion.

We consider a $C^{*}$ algebra $\mathcal{A}$ and a locally compact group $G$, which to simplify the writing we assume to be Abelian, acting on $\mathcal{A}$ such that for every $g\in G$, there is a linear map $\bar{g}: a\in \mathcal{A} \rightarrow a(g) \in \mathcal{A}$ with the properties,
    \begin{equation}
        ab(g)=a(g)b(g)\, \qquad \text{ and }
        a^{\dagger}(g)=a(g)^{\dagger}\, \qquad \text{ for } a,b \in \mathcal{A}
    \end{equation}
that preserves the norm of $a$ in $\mathcal{A}$, $|a|=|a(g)|$. In other words the map preserves the group structure of $\mathcal{A}$. It is also a linear map over complex numbers such that,
\begin{equation}
  (\alpha a+\beta b)(g)=\alpha a(g)+\beta b(g) \, \qquad \text{ for } \alpha,\beta \text{ complex numbers.}
\end{equation}

Note that this furnishes a representation of the group $G$ in the automorphism group of $\mathcal{A}$, that is,
\begin{align}
    [a(g_{1})](g_{2})&=a(g_{1}+g_{2})\\
    a(0)=a.
\end{align}
In addition, the function $a(g)$ is a continuous function of $g\in G$ in the norm topology of $\mathcal{A}$.

We define the covariance algebra $(\mathcal{A},G)$ as a set of all measurable functions, $A$, from $G$ to $\mathcal{A}$,
\begin{equation}
    A: g\in G\rightarrow A_{g}\in\mathcal{A}
\end{equation}
defined up to a measure zero and absolutely integrable set as long as,
\begin{equation}
    |A|_{1}=\int |A_{g}| dg <\infty,
\end{equation}
This means the function $A$ is not necessarily exactly everywhere defined and $dg$ is the Haar measure on $G$. Note that the element $A$ is not an operator in $\mathcal{A}$, rather a function which takes values in $\mathcal{A}$. The product of $A$ and $B$, elements of $(\mathcal{A},G)$, is defined as follows,
\begin{equation}
   (A.B)_{g}=\int A_{u}B_{g-u}(u)du 
\end{equation}
where $B_{g-u}(u)$ is the image under the action of the automorphism group by $u$, of the element $B_{g-u}$; while the element $B_{g-u}$ is just the image of $B$, an element of the covariance algebra at $g-u \in G$. We define the adjoint of $A$ as,
\begin{equation}
    A^{\dagger}_{g}=A_{-g}(g)^{\dagger}
\end{equation}
so that $|A^{\dagger}|=\int |A_{-g}(g)^{\dagger}|dg=\int |A_{-g}|dg=|A|$.

With these definitions of the adjoint and products of the elements of the covariance algebra, it can be shown that it forms a Banach $^{*}-$ algebra. In section 3 of \cite{1966CMaPh...3....1D} the representations of this algebra are shown to be in one to one correspondence with the covariant representation of $\mathcal{A}$. This rather unfamiliar form for the covariance algebra is discussed in the familiar form of crossed product algebra \cite{Witten:2021unn,Chandrasekaran:2022eqq} when $\mathcal{A}$ is a von Neumann algebra, in work by Takesaki \cite{Takesaki1973DualityFC}.

\bibliographystyle{JHEP}
\bibliography{ABh1}

\end{document}